\documentclass{article}
\usepackage[english]{babel}
\usepackage[T1]{fontenc}
\usepackage[ansinew]{inputenc}
\usepackage{graphicx}

\title{Subdiffusion in a system consisting of two different media separated by a thin membrane}

\author{Tadeusz Koszto{\l}owicz}

\date{\footnotesize{Institute of Physics, Jan Kochanowski University,\\
         ul. \'Swi\c{e}tokrzyska 15, 25-406 Kielce, Poland    \\  e-mail: tadeusz.kosztolowicz@ujk.edu.pl}}

\begin{document}
\maketitle

\begin{abstract}
We consider subdiffusion in a system which consists of two media separated by a thin membrane. The subdiffusion parameters may be different in each of the  medium. Using the new method presented in this paper we derive the probabilities (the Green's functions) describing a particle's random walk in the system. Within this method we firstly consider the particle's random walk in a system with both discrete time and space variables in which a particle can vanish due to reactions with constant probabilities $R_1$ and $R_2$ defined separately for each medium. Then, we move from discrete to continuous variables. The reactions included in the model play a supporting role. We link the reaction probabilities with the other subdiffusion parameters which characterize the media by means of the formulae presented in this paper. Calculating the generating functions for the difference equations describing the random walk in the composite membrane system with reactions, which depend explicitly on $R_1$ and $R_2$, we are able to correctly incorporate the subdiffusion parameters of both the media into the Green's functions. Finally, placing $R_1=R_2=0$ into the obtained functions we get the Green's functions for the composite membrane system without any reactions. From the obtained Green's functions, we derive the boundary conditions at the thin membrane. One of the boundary conditions contains the Riemann--Liouville fractional time derivatives, which show that the additional `memory effect' is created by a discontinuoity of the system. The second boundary condition demands continuity of a flux at the membrane. 

\end{abstract}

{\it keywords}: subdiffusion, diffusion in composite systems with a thin membrane, random walk model

\section{Introduction\label{SecI}}

In the many processes considered in biological, engineering or physical sciences normal diffusion or subdiffusion occurs in a system composed of two media, separated by a partially permeable thin membrane; in each part of the system different parameters characterizing diffusion may occur. 
As an example, we mention the transport of various substances (glucose, pyruvate, lactate, alanine) between blood and a cell \cite{kim}, diffusion of various substances (medications, cosmetics) through the skin \cite{schumm}, diffusion of various substances in the brain \cite{zhan},
diffusion between extracellular brain space and cells \cite{tao}, diffusion of large molecular drugs into the brain \cite{linninger}, nisin diffusion into agarose gel \cite{sebti,deleris}. A list of similar examples can be significantly expanded, see also the problems discussed in \cite{hobbie,luckey,hsieh}.

Subdiffusion is usually defined as a random walk process in which \cite{mk}
\begin{equation}\label{eq1}
\left\langle (\Delta x)^2\right\rangle=\frac{2D t^\alpha}{\Gamma(1+\alpha)}\;,
\end{equation} 
where $\left\langle (\Delta x)^2\right\rangle$ is the mean square displacement of a random walker, $\alpha$ is a subdiffusion parameter ($0<\alpha<1$), $D$ is a subdiffusion coefficient, measured in units $m^2/s^\alpha$, and $\Gamma$ denotes the Gamma function. The process of subdiffusion can occur in media in which particles' movement is greatly hindered due to the internal structure of a medium such as, for example, in gels \cite{mk1,kdm,kdm1}.
Subdiffusion is usually described by the following subdiffusion equation with the Riemann--Liouville fractional time derivative (here $0<\alpha<1$) \cite{mk}
\begin{equation}\label{eq2}
\frac{\partial C(x,t)}{\partial t}=D\frac{\partial^{1-\alpha}}{\partial t^{1-\alpha}}\frac{\partial^2 C(x,t)}{\partial x^2}\;.
\end{equation}
The Riemann--Liouville fractional derivative is defined as being valid for $\delta>0$ (here $k$ is a natural number which fulfils $k-1\leq \delta <k$)
\begin{equation}\label{eq3}
\frac{d^\delta f(t)}{dt^\delta}=\frac{1}{\Gamma(k-\delta)}\frac{d^k}{dt^k}\int_0^t{(t-t')^{k-\delta-1}f(t')dt'}\;.
\end{equation}
For $\alpha=1$ one obtains a normal diffusion equation.

We consider subdiffusion in a system which consists of two media separated by a thin partially permeable membrane. In each part of the system, bounded by the membrane, there may be different subdiffusion parameters. We assume that the system is homogeneous in a plane perpendicular to the $x$ axis, thus the system is effectively one--dimensional. The thin membrane is treated here as a partially permeable wall. A diffusing particle which tries to pass through the membrane can be stopped by the membrane with some probability or can jump through the membrane to another part of the system. The membrane is assumed to be so thin that the particle's random walk inside the membrane is not considered.
To solve the subdiffusion equation in a membrane system, two boundary conditions at the membrane are needed. However, until now, various boundary conditions which are not equivalent to one another have been assumed at the membrane, see for example \cite{zhang,grebenkov,kpre,k1,k2,k3,k4,kdl,dworecki,tk,tk1,korabel1,korabel2,korabel3,singh,kim1,ash,huang,adrover,abdekhodaie,taveira,cabrera1,cabrera2}. 

We present the method of deriving the probability $P(x,t;x_0)$ (the Green's function) of finding a particle at point $x$ at time $t$, $x_0$ is the initial position of a particle in the composite system with a thin membrane. 
We derive the Green's functions using a simple random walk model with both discrete time and spatial variables. Next we move to the continuous variables by means of the procedure presented in this paper. From the Green's functions we derive boundary conditions at a thin membrane.
In the paper \cite{tk} the method based on the random walk model on a lattice was used to determine the Green's functions for a system with a thin membrane in which the homogeneous medium is divided by a thin membrane into two parts. However, this method is not applicable to the system in which the subdiffusion parameters are different at each part of the system. In this paper we present a new method of deriving the Green's functions for the composite system with a thin membrane. 

A random walk model with a discrete time and space variable appears to be a useful tool in modeling anomalous or normal diffusion \cite{gillespie,ibe,ks,hk,chandrasekhar,hughes}. This model is particularly conducive in determining the Green's function for a system in which the uniformity of a system is disturbed at one point \cite{hughes,barber,montroll65,montroll64,weiss}; in this paper the point represents a membrane which is located perpendicularly to the $x$ axis in a one--dimensional system. In order to move from discrete to continuous time in a homogeneous system one assumes that the time which is needed for a particle to take its next step is ruled by the distribution $\omega(t)$. However, the situation is more complicated in a system which consists of two parts with different transport properties since the two distributions $\omega_1(t)$ and $\omega_2(t)$, which describe the time which a particle needs to take its next step in the regions $x<x_N$ and $x>x_N$ ($x_N$ denotes the position of a membrane), respectively, are required. 

In order to incorporate correctly these distributions into the Green's function obtained for a discrete time, it is desirable to know how many particle's steps are performed in each region. 
In this paper we will show the new method of calculating the Green's functions for a membrane system, which omits the difficulties described above and is relatively simple to use. This method is based on the `mathematical trick', which consists of considering the random walk of a particle $Q$ in a system with two kinds of reactions with static particles $Q_A$ and $Q_B$. The first reaction, $Q+Q_A\longrightarrow Q_A$, can occur with probability $R_1$ in the region $x<x_N$, the second reaction, $Q+Q_B\longrightarrow Q_B$, can occur with probability $R_2$, in the region $x>x_N$. Supposing that the particles $Q_A$ and $Q_B$ are distributed homogeneously in the appropriate parts of the system, the reaction probabilities do not depend on a space variable. Solving the difference equations describing the random walk with reactions (more specifically finding the generating functions for these equations) and connecting the functions $\omega_1(t)$ and $\omega_2(t)$ with the reaction probabilities $R_1$ and $R_2$, respectively, we finally obtain the Green's functions in which the functions $\omega_1(t)$ and $\omega_2(t)$ are correctly incorporate in the model. Following this, in order to obtain the Green's functions for the system without recations, we place $R_1=R_2=0$.  In the last step, we move from a discrete to continuous space variable. 

The organization of the paper is as follows. In Sec. \ref{SecII}, we present the model and we find the general form of the Green's functions for the system under consideration. Using the Green's functions, we derive the general form of boundary conditions at the membrane. In Sec. \ref{SecIII}, we derive the Laplace transforms of Green's functions. From these functions we derive the boundary conditions at the membrane in terms of the Laplace transform. In Sec. \ref{SecIV}, we find the Green's functions and boundary conditions in time domain for the following cases. The first one, when a particle starts its random walk in a `faster' medium, the second one in which a particle starts its random walk in a `slower' medium, and the third one in which both media have the same subdiffusion parameters but different subdiffusion coefficients. The `faster' medium is defined as a medium having a larger parameter $\alpha$. The properties of the Green's functions and boundary conditions at the membrane are also discussed. The final remarks are presented in Sec. \ref{SecV}. Some details of the calculations are presented in the two Appendixes.

\section{Model\label{SecII}}

We suppose that a thin membrane located at the point $x_N$ separates the region $x<x_N$ (in the following, denoted as $A$) with subdiffusion parameters $D_1$, $\alpha_1$ and the region $x>x_N$ (denoted as $B$) with the parameters $D_2$, $\alpha_2$, $0<\alpha_1,\alpha_2\leq 1$. 
The Green's function, a generating function, and a flux defined for $x< x_N$ and $m\leq N$ are all labelled by the index $A$, whereas the functions defined for $x> x_N$ and $m\geq N+1$ are labelled by the index $B$. 

\begin{figure}[!ht]
\centering
\includegraphics[height=2.5cm]{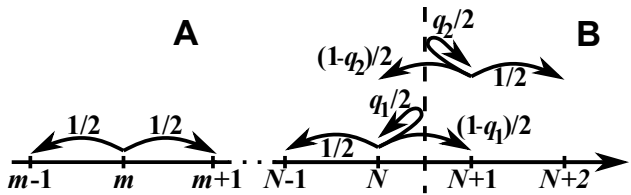}
\caption{A composite system with a thin membrane represented by the vertical dashed line, $A$ and $B$ denote the regions $m\leq N$ (with parameters $\alpha_1$ and $D_1$) and $m\geq N+1$ (with parameters $\alpha_2$ and $D_2$), respectively, a more detailed description is in the text.}\label{Fig1}
\end{figure}

\subsection{Discrete time and space variable\label{SecIIA}}

Firstly, we consider a particle's random walk in a system in which time $n$ and spatial variable $m$ are discrete; the thin membrane is located between the $N$ and $N+1$ site (see Fig. 1). The particle can be absorbed due to the reaction with probability $R_1$ in the region $m\leq N$, and with probability $R_2$ in the region $m\geq N+1$. In the following, $P_{i,n}(m;m_0)$ is the probability of finding the particle at site $m$, after $n$ steps, in the region $i=A,B$, and $m_0$ denotes the initial position of the particle.

A particle performs its single jump to the neighboring site only if the particle is not stopped by the wall with a certain probability. The particle which tries to pass through the membrane moves from the $N$ to $N+1$ site and can pass the membrane with probability $(1-q_1)/2$ or can be stopped by the membrane with probability $q_1/2$. When a particle is located at the $N+1$ site, then its jump to the $N$ site can be performed with probability $(1-q_2)/2$. The probability that a particle can be stopped by the wall equals $q_2/2$. In this paper we assume that $q_1,q_2\neq 0$.
The difference equations describing the random walk in a membrane system with reactions read as follows
\begin{eqnarray}
 \label{eq4}
 P_{A,n+1}(m;m_0)=\frac{1}{2}P_{A,n}(m-1;m_0)+\frac{1}{2}P_{A,n}(m+1;m_0)  \nonumber\\ 
-R_1 P_{A,n}(m;m_0),\;\;  m\leq N-1,
\end{eqnarray}
\begin{eqnarray}
 \label{eq5}
P_{A,n+1}(N;m_0)=\frac{1}{2}P_{A,n}(N-1;m_0)+\frac{1-q_2}{2}P_{B,n}(N+1;m_0)\nonumber\\
+\frac{q_1}{2}P_{A,n}(N;m_0)-R_1 P_{A,n}(N;m_0)\;,
\end{eqnarray}
\begin{eqnarray}
 \label{eq6}
P_{B,n+1}(N+1;m_0)=\frac{1-q_1}{2}P_{A,n}(N;m_0)+\frac{1}{2}P_{B,n}(N+2;m_0)\nonumber\\ 
+\frac{q_2}{2}P_{B,n}(N+1;m_0)-R_2 P_{B,n}(N+1;m_0)\;,
 \end{eqnarray}
\begin{eqnarray}
 \label{eq7}
P_{B,n+1}(m;m_0)=\frac{1}{2}P_{B,n}(m-1;m_0)+\frac{1}{2}P_{B,n}(m+1;m_0)\nonumber\\
-R_2 P_{B,n}(m;m_0),\;\; m\geq N+2,
\end{eqnarray}
the initial condition is 
\begin{equation}\label{eq8}
P_{i,0}(m;m_0)=\delta_{m,m_0}\;,
\end{equation} 
$i=A,B$. In the following, we assume that $m_0\leq N$.
The generating functions are defined as
\begin{equation}\label{eq9}
  S_i(m,z;m_0)=\sum_{n=0}^{\infty}z^nP_{i,n}(m,m_0)\;,
\end{equation}
$i=A,B$. 
After calculations, we obtain (the details of the calculations are presented in Appendix I)
\begin{equation}
  \label{eq10} 
S_A(m,z;m_0)=\frac{[\eta_{1}(z)]^{|m-m_0|}}{\sqrt{(1+zR_1)^2-z^2}}+\Lambda_A(z)\frac{[\eta_{1}(z)]^{2N-m-m_0}}{\sqrt{(1+zR_1)^2-z^2}}\;,
\end{equation}
\begin{equation}
  \label{eq11} 
  S_B(m,z;m_0)=\frac{[\eta_{1}(z)]^{N-m_0}[\eta_{2}(z)]^{m-N-1}}{\sqrt{(1+zR_2)^2-z^2}}\Lambda_B(z)\;,
\end{equation}
where
\begin{equation}
  \label{eq12}
\Lambda_A(z)=\frac{\Big(\frac{1}{\eta_{2}(z)}-q_2\Big)\Big(q_1-\eta_{1}(z)\Big)+(1-q_1)(1-q_2)}{\Big(\frac{1}{\eta_{1}(z)}-q_1\Big)\Big(\frac{1}{\eta_{2}(z)}-q_2\Big)-(1-q_1)(1-q_2)}\;,
\end{equation}
\begin{equation}
  \label{eq13}
\Lambda_B(z)=\frac{(1-q_1)\Big(\frac{1}{\eta_{2}(z)}-\eta_{2}(z)\Big)}{\Big(\frac{1}{\eta_{1}(z)}-q_1\Big)\Big(\frac{1}{\eta_{2}(z)}-q_2\Big)-(1-q_1)(1-q_2)}\;,
\end{equation}
\begin{equation}\label{eq14}
\eta_{i}(z)=\frac{1+R_iz-\sqrt{(1+R_iz)^2-z^2}}{z}\;.
\end{equation}

\subsection{From discrete to continuous time\label{SecIIB}}

Let $\omega_1(t)$ denotes the probability distribution of time which is needed to the performing particle's step in the region $m\leq N$ and $\omega_2(t)$ the similar probability distribution defined for the region $m\geq N+1$. In the system in which $\omega_1(t)\equiv\omega_2(t)\equiv\omega(t)$ in terms of the Laplace transform, $\mathcal{L}[f(t)]\equiv \hat{f}(s)=\int_0^\infty {\rm e}^{-st}f(t)dt$, the Green's function for continuous time and a discrete spatial variable are expressed by the following general formula \cite{montroll65} $\hat{P}(m,s;m_0)=\hat{U}(s)S(m,\hat{\omega}(s);m_0)$, where $\hat{U}(s)=(1-\hat{\omega}(s))/s$
is the Laplace transform of function $U(t)=1-\int_0^t \omega(t')dt'$, which means that a particle has not performed any step over the time interval $(0,t)$. For the system under consideration we have
\begin{equation}\label{eq15}
\hat{P}_{A}(m,s;m_0)=\hat{U}_{1}(s)S_{A}(m,\{\hat{\omega}_1(s),\hat{\omega}_2(s)\};m_0)\;,
\end{equation}
\begin{equation}\label{eq15b}
\hat{P}_{B}(m,s;m_0)=\hat{U}_{2}(s)S_{B}(m,\{\hat{\omega}_1(s),\hat{\omega}_2(s)\};m_0)\;,
\end{equation}
where 
\begin{equation}\label{eq15a}
\hat{U}_i(s)=\frac{1-\hat{\omega}_i(s)}{s}\;,
\end{equation}
$i=1,2$, the symbol $\{\hat{\omega}_1(s),\hat{\omega}_2(s)\}$ denotes that both of the functions $\hat{\omega}_1(s)$ or $\hat{\omega}_2(s)$ can be involved into the functions $S_A$ and $S_B$ instead of the variable $z$. In order to find the rule according which such replacement should be performed we consider a distribution of the first passage time. Let $F_n(m;m_0)$ denotes a probability that the random walker, which starts to its movement form a point $m_0$, arrive to point $m$ at first time in $n$th step. The generating function, defined as $S_F(m,z;m_0)=\sum_{n=1}^\infty z^n F_n(m;m_0)$, fulfils the following equation 
\begin{equation}\label{eq101}
S_F(m,z;m_0)=\frac{S(m,z;m_0)-\delta_{m,m_0}}{S(m,z;m)}\;. 
\end{equation}
Assuming $m_0<N$, from Eqs. (\ref{eq10}), (\ref{eq12}), and (\ref{eq101}) we get 
\begin{equation}\label{eq102}
S_F(N,z;m_0)=\eta_{1}^{N-m_0}(z)\;.
\end{equation}
Since all steps from $m_0$ to $N$ are performed in the region $A$, the first passage time distribution for continuous time reads in terms of the Laplace transform $\hat{F}(N,s;m_0)=\sum_{n=1}^\infty \hat{\omega}_1^n(s)F_n(N+1;m_0)\equiv S_F(N+1,\hat{\omega}_1;m_0)$. Finally we obtain
\begin{equation}\label{eq103}
\hat{F}(N,s;m_0)=\eta_1^{N-m_0}(\hat{\omega}_1(s))\;.
\end{equation}
Eq. (\ref{eq103}) shows that the function $\eta_1$ dependents on the function $\hat{\omega}_1$ only. In a similar way one can show that the function $\eta_2$ dependents on the function $\hat{\omega}_2$ only. Thus, the replacement of $z$ by $\hat{\omega}_1$ and $\hat{\omega}_2$ in Eqs. (\ref{eq10})--(\ref{eq15b}) should be performed by the following rules
\begin{equation}\label{eq104}
\eta_i(z)\rightarrow \eta_i(\hat{\omega}_i(s))\;,
\end{equation}
and
\begin{equation}\label{eq105}
\sqrt{(1+R_iz)^2-z^2}\rightarrow \sqrt{(1+R_i\omega_i(s))^2-\omega_i^2(s)}\;,
\end{equation}
$i=1,2$; Eq. (\ref{eq105}) is derived from Eqs. (\ref{eq14}) and (\ref{eq104}). We note that the replacement of $z$ by $\hat{\omega}_1$ or $\hat{\omega}_2$ in the denominator of $S_A$ and $S_B$, Eqs. (\ref{eq10}) and (\ref{eq11}), would not be obvious if $R_1=R_2=0$. This remark shows that the use of discrete equations describing particle's random walk in a system with reactions allows for an unambiguous derivation of Green's functions for the composite system.

Let us return to a membrane system without any reaction. Therefore, we place $R_1=R_2=0$ in the above formulae. Then, Eqs. (\ref{eq10})--(\ref{eq15a}), (\ref{eq104}), and (\ref{eq105}) provide
\begin{eqnarray}
  \label{eq26} 
\hat{P}_A(m,s;m_0)=\frac{1-\hat{\omega}_1(s)}{s\sqrt{1-\hat{\omega}_1^2(s)}}\left([\eta_1(\hat{\omega}_1(s))]^{|m-m_0|}\right.\nonumber\\
\left.+\tilde{\Lambda}_A(s)[\eta_1(\hat{\omega}_1(s))]^{2N-m-m_0}\right)\;,
\end{eqnarray}
\begin{equation}
  \label{eq27} 
\hat{P}_B(m,s;m_0)=\frac{1-\hat{\omega}_2(s)}{s\sqrt{1-\hat{\omega}_2^2(s)}}\tilde{\Lambda}_B(s)[\eta_1(\hat{\omega}_1(s))]^{N-m_0}[\eta_2(\hat{\omega}_2(s))]^{m-N-1}\;, 
\end{equation}
where
\begin{equation}
  \label{eq28} \tilde{\Lambda}_A(s)=\frac{\Big(\frac{1}{\eta_2(\hat{\omega}_2(s))}-q_2\Big)\Big(q_1-\eta_1(\hat{\omega}_1(s))\Big)+(1-q_1)(1-q_2)}{\Big(\frac{1}{\eta_1(\hat{\omega}_1(s))}-q_1\Big)\Big(\frac{1}{\eta_2(\hat{\omega}_2(s))}-q_2\Big)-(1-q_1)(1-q_2)}\;,
\end{equation}
\begin{equation}
  \label{eq29}
\tilde{\Lambda}_B(s)=\frac{(1-q_1)\Big(\frac{1}{\eta_2(\hat{\omega}_2(s))}-\eta_2(\hat{\omega}_2(s))\Big)}{\Big(\frac{1}{\eta_1(\hat{\omega}_1(s))}-q_1\Big)\Big(\frac{1}{\eta_2(\hat{\omega}_2(s))}-q_2\Big)-(1-q_1)(1-q_2)}\;,
\end{equation}
\begin{equation}\label{eq29a}
\eta_{i}(\hat{\omega}_i(s))=\frac{1-\sqrt{1-\hat{\omega}_i^2(s)}}{\hat{\omega}_i(s)}\;.
\end{equation}

\subsection{From a discrete to continuous space variable\label{SecIIC}}

Supposing that $\epsilon$ denotes the distance between discrete sites, and that 
\begin{equation}\label{eq30}
x=\epsilon m\;,\; x_0=\epsilon m_0\;,\;x_N=\epsilon N\;,
\end{equation} 
\begin{equation}\label{eq30a}
P(x,t;x_0)= \frac{P(m,t;m_0)}{\epsilon}\;,
\end{equation}
we transfer the Green's functions from a discrete to continuous space variable presuming $\epsilon$ to constitute only small values. 

In the continuous time random walk approach, the considerations are performed supposing that $s$ is of small value which corresponds to the limit of long time due to the Tauberian theorem. For small values of $s$, we may assume
\begin{equation}\label{eq31}
\hat{\omega}_i(s)=1-\tau_{\alpha_i} s^{\alpha_i}\;,
\end{equation}
$i=1,2$, where $\tau_{\alpha_i}$ is a parameter, measured in units of time raised to the power $\alpha_i$, which ensures the dimensionless form of the last term occurring in Eq. (\ref{eq31}). The parameter $\tau_{\alpha_i}$ is related to the distance between discrete sites, by the subdiffusion coefficient $D_i$ which is defined as \cite{tk,tk1}
\begin{equation}\label{eq32}
D_i=\frac{\epsilon^2}{2\tau_{\alpha_i}}\;. 
\end{equation}
However, in order to derive the subdiffusion equation, the limit of small values of $s$ is not necessary, since, for 
\begin{equation}\label{eq33}
\hat{\omega}_i(s)=\frac{1}{1+\tau_{\alpha_i} s^{\alpha_i}}
\end{equation}
the random walk model provides a subdiffusion equation with the Riemann--Liouville fractional time derivative (\ref{eq2}) without any restriction of the $s$ variable (see the discussion presented in \cite{kl2014}). 
From Eqs. (\ref{eq31})--(\ref{eq33}) we obtain for small values of $\epsilon$
\begin{equation}\label{eq34}
\hat{\omega}_i(s)=1-\frac{\epsilon^2}{2D_i} s^{\alpha_i}\;.
\end{equation}
We note that the moving form a discrete to continuous spatial variable cannot be regarded as a strict limit $\epsilon\longrightarrow 0$, because then the parameter $\tau_{\alpha_i}$, due to Eq. (\ref{eq32}), will also go to zero and the function $\omega_i(t)$ will lose its probabilistic interpretation. Therefore, performing the transition to a continuous spatial variable, the parameter $\epsilon$ is treated as non-zero, but of a small value.

The transition from a discrete to continuous space variable in Eqs. (\ref{eq26}) and (\ref{eq27}) is done separately for the function type $[\eta(\hat{\omega}_i(s))]^{|m|}$ and for the function $\tilde{\Lambda}_i(s)$.
We note that Eqs. (\ref{eq29a}) and (\ref{eq34}) provide for small values of $\epsilon$
\begin{equation}\label{eq35}
\eta_i\left(\hat{\omega}_i(s)\right)=1-\epsilon\sqrt{\frac{s^{\alpha_i}}{D_i}}\;,
\end{equation}
$i=1,2$. 
Taking into account Eqs. (\ref{eq30}) and (\ref{eq35}), we suppose that the following relation is satisfied for all functions occurring in Eqs. (\ref{eq26}) and (\ref{eq27}) 
\begin{equation}\label{eq36}
\left[\eta_i\left(\hat{\omega}_i(s)\right)\right]^k\approx\left(1-\epsilon\frac{s^{\alpha_i/2}}{\sqrt{D_i}}\right)^{|x_b|/\epsilon}\approx{\rm e}^{-\frac{|x_b| s^{\alpha_i}}{\sqrt{D_i}}}\;,
\end{equation}
where $i=1,2$ and $x_b=\epsilon k$. The `limit of small values of $\epsilon$' can be defined as of such a small value of $\epsilon$, that Eq. (\ref{eq36}) is fulfilled for all the functions of this type occurring in Eqs. (\ref{eq26}) and (\ref{eq27}).

For very small $\epsilon$, the frequency of jumps performed by a particle takes an anomalously large values \cite{ks}. A very large numbers of attempts to pass through the partially permeable membrane made over time interval $(0,t)$ for arbitrarily small values of $t$, means that the particle passes through the membrane with a probability equal to one. In this situation, the membrane loses its selective property. To avoid this, we suppose that the probabilities $q_1$ and $q_2$ are functions of $\epsilon$, in such a way that $q_1(0)=q_2(0)=1$ \cite{tk,tk1}. The calculation presented in Appendix II shows that
\begin{equation}\label{eq37}
q_1(\epsilon)=1-\frac{\epsilon}{\gamma_1}\;,\;q_2(\epsilon)=1-\frac{\epsilon}{\gamma_2}\;,
\end{equation}
where $\gamma_1$ and $\gamma_2$ are reflection membrane coefficients defined for the continuous system. 

The further considerations are performed in the limit of small values of $\epsilon$, as well as in the limit of small values of $s$. The limit of small values of $\epsilon$ means that this parameter will be absent in the obtained formulae, whereas the limit of small values of $s$ means that only the leading terms, with respect to this variable, will be present in considered functions.

\section{Laplace transforms of Green's functions and boundary conditions\label{SecIII}}

Using Eqs. (\ref{eq26}), (\ref{eq27}), (\ref{eq30}), (\ref{eq30a}), (\ref{eq32}), (\ref{eq34}), (\ref{eq36}) we obtain the Laplace transform of the Green's functions in the limit of small values of $\epsilon$
\begin{equation}
  \label{eq40}
\hat{P}_A(x,s;x_0)=\frac{s^{-1+\alpha_1/2}}{2\sqrt{D_1}}\left[{\rm e}^{-\frac{|x-x_0|s^{\alpha_1/2}}{\sqrt{D_1}}}+\tilde{\Lambda}_A(s){\rm e}^{-\frac{(2x_N-x-x_0)s^{\alpha_1/2}}{\sqrt{D_1}}}\right]\;,
\end{equation}
\begin{equation}
  \label{eq41}
\hat{P}_B(x,s;x_0)=\frac{s^{-1+\alpha_2/2}}{2\sqrt{D_2}}\;\tilde{\Lambda}_B(s) {\rm e}^{-\frac{(x_N-x_0)s^{\alpha_1/2}}{\sqrt{D_1}}}{\rm e}^{-\frac{(x-x_N)s^{\alpha_2/2}}{\sqrt{D_2}}}\;,
\end{equation}
where 
\begin{equation}\label{eq38}
\tilde{\Lambda}_A(s)=\frac{\tilde{\gamma}_1\sqrt{s^{\alpha_1}}-\tilde{\gamma}_2\sqrt{s^{\alpha_2}}+\tilde{\gamma}_1\tilde{\gamma}_2\sqrt{s^{\alpha_1+\alpha_2}}}{\tilde{\gamma}_1\sqrt{s^{\alpha_1}}+\tilde{\gamma}_2\sqrt{s^{\alpha_2}}+\tilde{\gamma}_1\tilde{\gamma}_2\sqrt{s^{\alpha_1+\alpha_2}}}\;,
\end{equation}
\begin{equation}\label{eq39}
\tilde{\Lambda}_B(s)=\frac{2\tilde{\gamma}_2\sqrt{s^{\alpha_2}}}{\tilde{\gamma}_1\sqrt{s^{\alpha_1}}+\tilde{\gamma}_2\sqrt{s^{\alpha_2}}+\tilde{\gamma}_1\tilde{\gamma}_2\sqrt{s^{\alpha_1+\alpha_2}}}\;,
\end{equation}
\begin{equation}\label{eq39a}
\tilde{\gamma}_i=\frac{\gamma_i}{\sqrt{D_i}}\;, 
\end{equation}
$i=1,2$. We note that Eqs. (\ref{eq38}) and (\ref{eq39}) provide $\tilde{\Lambda}_A(s)+\tilde{\Lambda}_B(s)=1$.

Functions (\ref{eq40}) and (\ref{eq41}) fulfil the following boundary conditions 
\begin{equation}\label{eq42}
\hat{P}_A(x_N,s;x_0)=\hat{\Phi}(s)\hat{P}_B(x_N,s;x_0)\;,
\end{equation}
where
\begin{equation}\label{eq43}
\hat{\Phi}(s)=\sqrt{\frac{D_2}{D_1}}s^{(\alpha_1-\alpha_2)/2}\frac{1+\tilde{\Lambda}_A(s)}{\tilde{\Lambda}_B(s)}
=\sqrt{\frac{D_1}{D_2}}\left[\frac{\tilde{\gamma}_1}{\tilde{\gamma}_2}s^{\alpha_1-\alpha_2}+\tilde{\gamma}_1s^{\alpha_1-\alpha_2/2}\right]\;.
\end{equation}
The second boundary condition concerns the fluxes flowing through the membrane. The subdiffusive flux $J$ is defined as
\begin{equation}\label{eq44}
J(x,t;x_0)=-D\frac{\partial^{1-\alpha}}{\partial t^{1-\alpha}}\frac{\partial}{\partial x}P(x,t;x_0)\;; 
\end{equation}
combining the above equation with the continuity equation $\partial P/\partial t=-\partial J/\partial x$, one obtains the subdiffusion equation (\ref{eq2}).
For $0<\delta<1$ and for a bounded function $f$, the Laplace transform of the Riemann--Liouville fractional derivative Eq. (\ref{eq3}) reads
\begin{equation}\label{eq45}
\mathcal{L}\left[\frac{d^\delta}{d t^\delta}f(t)\right]=s^\delta\hat{f}(s)\;.
\end{equation}
From Eqs. (\ref{eq44}) and (\ref{eq45}) we obtain
\begin{eqnarray}
\hat{J}_A(x,s;x_0)&=&-D_1 s^{1-\alpha_1}\frac{\partial \hat{P}_A(x,s;x_0)}{\partial x}\;,\label{eq46}\\
\hat{J}_B(x,s;x_0)&=&-D_2 s^{1-\alpha_2}\frac{\partial \hat{P}_B(x,s;x_0)}{\partial x}\;.\label{eq47}
\end{eqnarray}
Equations (\ref{eq40}), (\ref{eq41}), (\ref{eq46}) and (\ref{eq47}) provide the second boundary condition at the membrane
\begin{equation}\label{eq48}
\hat{J}_A(x_N,s;x_0)=\hat{J}_B(x_N,s;x_0)\;.
\end{equation}
Thus, the flux is continuous at the membrane
\begin{equation}\label{eq49}
J_A(x_N,t;x_0)=J_B(x_N,t;x_0)\;.
\end{equation}

\section{Green's functions and boundary conditions in time domain\label{SecIV}}

The frequency of a particle's jumps is greater for a higher parameter $\alpha$ then the movement of a particle is `faster'. For the case of $\alpha_1\neq\alpha_2$, we call the region with larger parameter $\alpha$ a `faster' region and the other region a `slower' region. Below, we consider subdiffusion in a composite system with a thin membrane for three qualitatively different cases. In the first case, a particle starts its random walk in the `faster' region, in the second one, it starts in the `slower' region and finally we consider subdiffusion in a system in which the parameter $\alpha$ is the same in both regions, but subdiffusion coefficients $D_1$ and $D_2$ may be different. 

The following considerations will be performed in the limit of small values of parameter $s$.
Functions $\tilde{\Lambda}_A(s)$ and $\tilde{\Lambda}_B(s)$ play a key role in the further considerations because they contain the parameters controlling membrane permeability $\tilde{\gamma}_1$ and $\tilde{\gamma}_2$. The approximation of functions $\tilde{\Lambda}_A(s)$ and $\tilde{\Lambda}_B(s)$, for small values of  $s$, provides the Green's functions which are qualitatively different for the cases $\alpha_1>\alpha_2$, $\alpha_1<\alpha_2$, and $\alpha_1=\alpha_2$. 
In order to obtain approximate forms of these functions we take two leading terms of $\tilde{\Lambda}_A(s)$ with respect to $s$, and next we use the relation $\tilde{\Lambda}_B(s)=1-\tilde{\Lambda}_A(s)$. 

In the following the inverse Laplace transform of the Green's functions will be calculated with the formula \cite{koszt}
\begin{equation}\label{eq58}
\mathcal{L}^{-1}\left[s^\nu {\rm e}^{-as^\beta}\right]\equiv f_{\nu,\beta}(t;a)
=\frac{1}{t^{\nu+1}}\sum_{k=0}^\infty{\frac{1}{k!\Gamma(-k\beta-\nu)}\left(-\frac{a}{t^\beta}\right)^k}\;,
\end{equation}
$a,\beta>0$; the function $f_{\nu,\beta}$ is a special case of the H-Fox function.
To roughly evaluate the time over which the approximation of small values of $s$ is valid, we use Eq. (\ref{eq58}). We note that the functions $s^\nu {\rm e}^{-as^\beta}$ and $f_{\nu,\beta}(t;a)$ can be approximated by their first leading terms under conditions $as^\beta\ll 1$ and $a/(\Gamma(-\beta-\nu)t^\beta)\ll 1/\Gamma(-\nu)$, respectively. Placing $\nu=-1$, we deduce that the condition $s\ll 1/a^{1/\beta}$ causes $t\gg (a/\Gamma(1-\beta))^{1/\beta}$, $a>0$. This condition can also be obtained through the strong Tauberian theorem \cite{hughes}.
We stress that the above condition should be treated as a rough estimation of the `long time limit' which corresponds to the `limit of small values of $s$'. Further calculations will be carry out on the assumption that $x_0<x_N$.

\subsection{From faster to slower region, $\alpha_1>\alpha_2$\label{SecIVA}}

Let $\alpha_1>\alpha_2$. We suppose that $\tilde{\gamma}_1\neq\tilde{\gamma}_2$. As we argue later in Sec. \ref{SecV}, in practice, this condition does not reduce the generality of the obtained results.
In order to carry out calculations in a clearer manner we transform Eqs. (\ref{eq38}) and (\ref{eq39}) into the form
\begin{equation}\label{eq50}
\tilde{\Lambda}_A(s)=\frac{-1+(\tilde{\gamma}_1/\tilde{\gamma}_2)s^{(\alpha_1-\alpha_2)/2}+\tilde{\gamma}_1 s^{\alpha_1/2}}{1+(\tilde{\gamma}_1/\tilde{\gamma}_2)s^{(\alpha_1-\alpha_2)/2}+\tilde{\gamma}_1 s^{\alpha_1/2}}\;,
\end{equation}
\begin{equation}\label{eq51}
\tilde{\Lambda}_B(s)=\frac{2}{1+(\tilde{\gamma}_1/\tilde{\gamma}_2)s^{(\alpha_1-\alpha_2)/2}+\tilde{\gamma}_1 s^{\alpha_1/2}}\;.		
\end{equation}
The Laplace transform of boundary condition (\ref{eq42}) reads
\begin{equation}\label{eq52}
\hat{P}_A(x_N,s;x_0)=\sqrt{\frac{D_1}{D_2}}\left[\frac{\tilde{\gamma}_1}{\tilde{\gamma}_2}s^{\alpha_1-\alpha_2}+\tilde{\gamma}_1s^{\alpha_1-\alpha_2/2}\right]\hat{P}_B(x_N,s;x_0)\;.
\end{equation}

Due to Eq. (\ref{eq45}), the boundary condition in the time domain reads
\begin{eqnarray}\label{eq53}
  P_A(x_N,t;x_0)=\sqrt{\frac{D_2}{D_1}}\frac{\tilde{\gamma}_1}{\tilde{\gamma}_2}\frac{\partial^{\alpha_1-\alpha_2}P_B(x_N,t;x_0)}{\partial t^{\alpha_1-\alpha_2}}\nonumber\\
  +\sqrt{\frac{D_2}{D_1}}\tilde{\gamma}_1\frac{\partial^{\alpha_1-\alpha_2/2}P_B(x_N,t;x_0)}{\partial t^{\alpha_1-\alpha_2/2}}\;.
\end{eqnarray}
For $s\ll \left(1/\tilde{\gamma}_2\right)^{2/\alpha_2}$ Eq. (\ref{eq52}) can be approximated as follows
\begin{equation}\label{eq54}
\hat{P}_A(x_N,s;x_0)=\sqrt{\frac{D_2}{D_1}}\frac{\tilde{\gamma}_1}{\tilde{\gamma}_2}s^{\alpha_1-\alpha_2}\hat{P}_B(x_N,s;x_0)\;.
\end{equation}
Thus, for $t\gg \left(\tilde{\gamma}_2/\Gamma(1-\alpha_2/2)\right)^{2/\alpha_2}$ the approximate form of the boundary condition at the membrane reads
\begin{eqnarray}\label{eq55}
P_A(x_N,t;x_0)=\frac{\sqrt{D_2}\tilde{\gamma}_1}{\sqrt{D_1}\tilde{\gamma}_2}\frac{\partial^{\alpha_1-\alpha_2}P_B(x_N,t;x_0)}{\partial t^{\alpha_1-\alpha_2}}\;.
\end{eqnarray}
The Laplace transforms of the Green's functions (\ref{eq40}) and (\ref{eq41}) with $\tilde{\Lambda}_A(s)$ and $\tilde{\Lambda}_B(s)$ given by Eqs. (\ref{eq50}) and (\ref{eq51}), respectively, take on rather complicated forms for which the inverse Laplace transform can be calculated in the limit of small values of $s$. 
Assuming additional condition $s\ll \left(\tilde{\gamma}_1/\tilde{\gamma}_2\right)^{2/(\alpha_1-\alpha_2)}$,
we obtain the following approximate form of the functions (\ref{eq50}) and (\ref{eq51})
$\tilde{\Lambda}_A(s)=-1+2(\tilde{\gamma}_1/\tilde{\gamma}_2)s^{(\alpha_1-\alpha_2)/2}$,
$\tilde{\Lambda}_B(s)=2\left(1-(\tilde{\gamma}_1/\tilde{\gamma}_2)s^{(\alpha_1-\alpha_2)/2}\right)$,
which together with Eqs. (\ref{eq40}) and (\ref{eq41}) give
\begin{eqnarray}\label{eq56}
\hat{P}_A(x,s;x_0)=\frac{s^{-1+\alpha_1/2}}{2\sqrt{D_1}}\Bigg[{\rm e}^{-\frac{|x-x_0|s^{\alpha_1/2}}{\sqrt{D_1}}}\\
-\left(1-\frac{2\tilde{\gamma}_1 s^{(\alpha_1-\alpha_2)/2}}{\tilde{\gamma}_2}\right){\rm e}^{-\frac{(2x_N-x-x_0)s^{\alpha_1/2}}{\sqrt{D_1}}}\Bigg]\nonumber\;,
\end{eqnarray}
\begin{equation}\label{eq57}
\hat{P}_B(x,s;x_0)=\left[\frac{s^{-1+\alpha_2/2}}{\sqrt{D_2}}-\frac{\tilde{\gamma}_1 s^{-1+\alpha_1/2}}{\tilde{\gamma}_2 \sqrt{D_2}}\right]{\rm e}^{-\frac{(x_N-x_0)s^{\alpha_1/2}}{\sqrt{D_1}}}{\rm e}^{-\frac{(x-x_N)s^{\alpha_2/2}}{\sqrt{D_2}}}\;.
\end{equation}
We expand an exponential function in Eq. (\ref{eq57}) to a power series, ${\rm e}^u=\sum_{n=0}^\infty u^n/n!$. For this expansion, we have chosen to use only one of the two exponential functions in Eq.~(\ref{eq57}), which will reach $1$ more rapidly for $s\longrightarrow 0$ (i.e. a function with larger values of the parameter $\alpha_i$ which controlls the exponent of the function).

For $t\gg {\rm max}[t_1,t_2]$, where $t_1=(\tilde{\gamma}_2/\Gamma(1-\alpha_2/2))^{2/\alpha_2}$, $t_2=\left(\tilde{\gamma}_1/\tilde{\gamma}_2\Gamma(1-(\alpha_1-\alpha_2)/2)\right)^{2/(\alpha_1-\alpha_2)}$, due to Eq. (\ref{eq58}), the Green's functions read
\begin{eqnarray}\label{eq59}
P_A(x,t;x_0)=\frac{1}{2\sqrt{D_1}}\Bigg[f_{-1+\alpha_1/2,\alpha_1/2}\left(t;\frac{|x-x_0|}{\sqrt{D_1}}\right)\\
-f_{-1+\alpha_1/2,\alpha_1/2}\left(t;\frac{2x_N-x-x_0}{\sqrt{D_1}}\right)\Bigg]\nonumber\\
+\frac{\tilde{\gamma}_1}{\sqrt{D_1}\tilde{\gamma}_2}f_{-1+\alpha_1-\alpha_2/2,\alpha_1/2}\left(t;\frac{2x_N-x-x_0}{\sqrt{D_1}}\right)\;,\nonumber
\end{eqnarray}
\begin{eqnarray}\label{eq60}
P_B(x,t;x_0)=\frac{1}{\sqrt{D_2}}\sum_{n=0}^\infty\frac{1}{n!}\left(\frac{x_0-x_N}{\sqrt{D_1}}\right)^n\\
\times\Bigg[f_{-1+\alpha_2/2+n\alpha_1/2,\alpha_2/2}\left(t;\frac{x-x_N}{\sqrt{D_2}}\right)\nonumber\\
-\frac{\tilde{\gamma}_1}{\tilde{\gamma}_2}f_{-1+(n+1)\alpha_1/2,\alpha_2/2}\left(t;\frac{x-x_N}{\sqrt{D_2}}\right)\Bigg]\;.\nonumber
\end{eqnarray}

\begin{figure}[!ht]
\centering
\includegraphics[height=6.0cm]{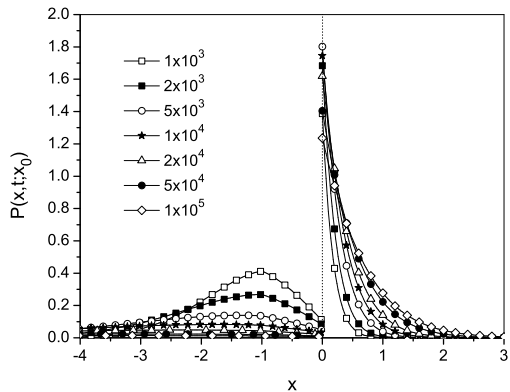}
\caption{Plots of the Green's functions (\ref{eq59}) and (\ref{eq60}) for various times given in the legend, $\alpha_1=0.9$, $\alpha_2=0.5$. The values of the other parameters are  $D_1=D_2=0.001$, $\tilde{\gamma}_1=0.06$, $\tilde{\gamma}_2=0.05$, $x_N=0$, $x_0=-1$, all quantities are given in arbitrarily chosen units.}\label{Fig2}
\end{figure}

\begin{figure}[!ht]
\centering
\includegraphics[height=6.0cm]{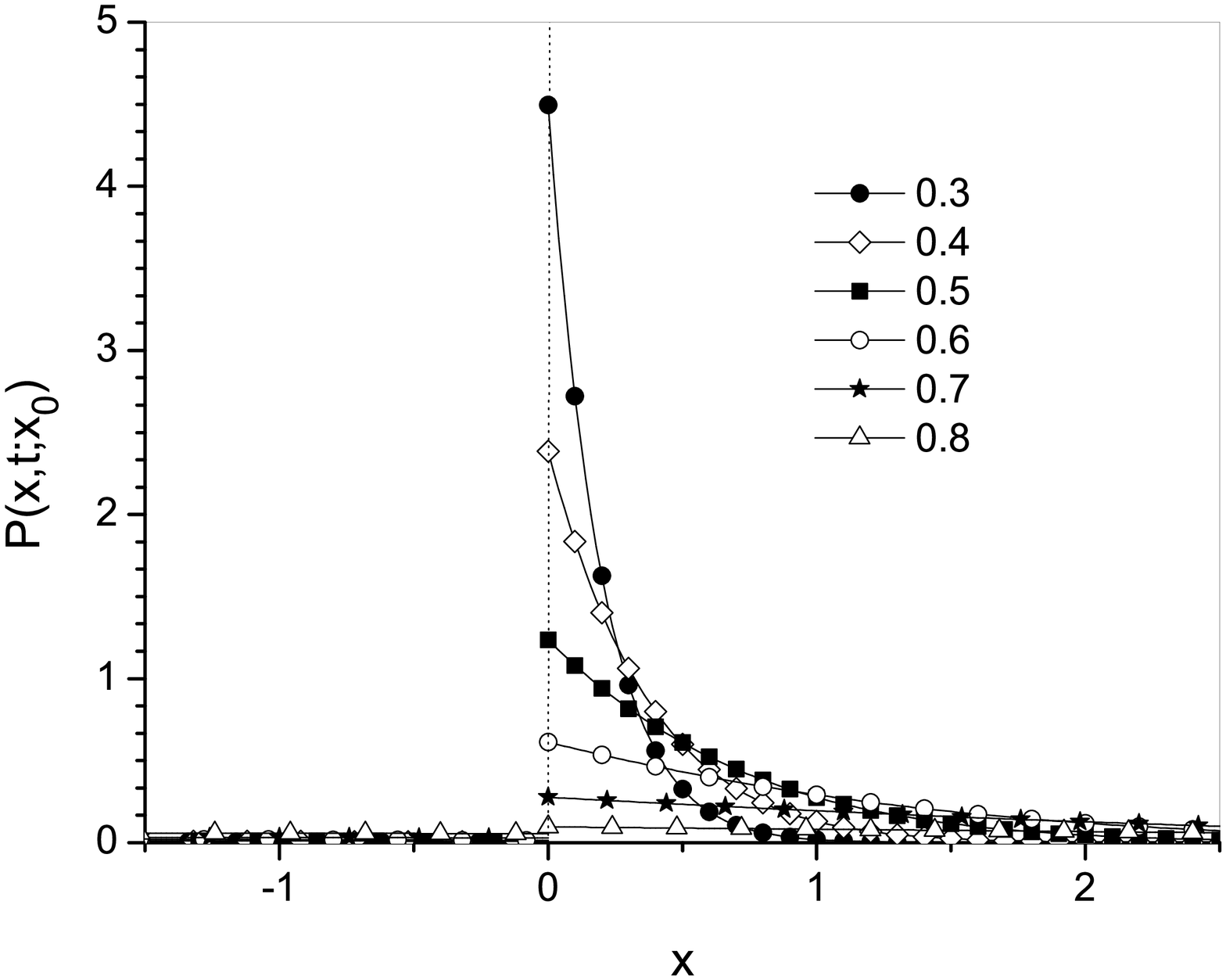}
\caption{The plots ot the Green's functions (\ref{eq59}) and (\ref{eq60}) for various $\alpha_2$ which are given in the legend, $\alpha_1=0.9$, $t=10^5$, the other parameters are the same as in Fig. \ref{Fig2}.}\label{Fig3}
\end{figure}

\begin{figure}[!ht]
\centering
\includegraphics[height=6.0cm]{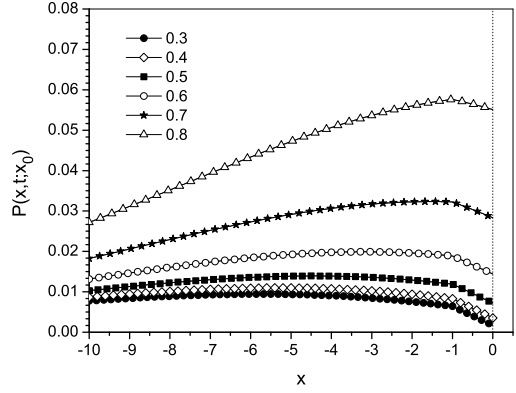}
\caption{The fragment of Fig. \ref{Fig3} done in the other scale.}\label{Fig4}
\end{figure}

\begin{figure}[!ht]
\centering
\includegraphics[height=6.0cm]{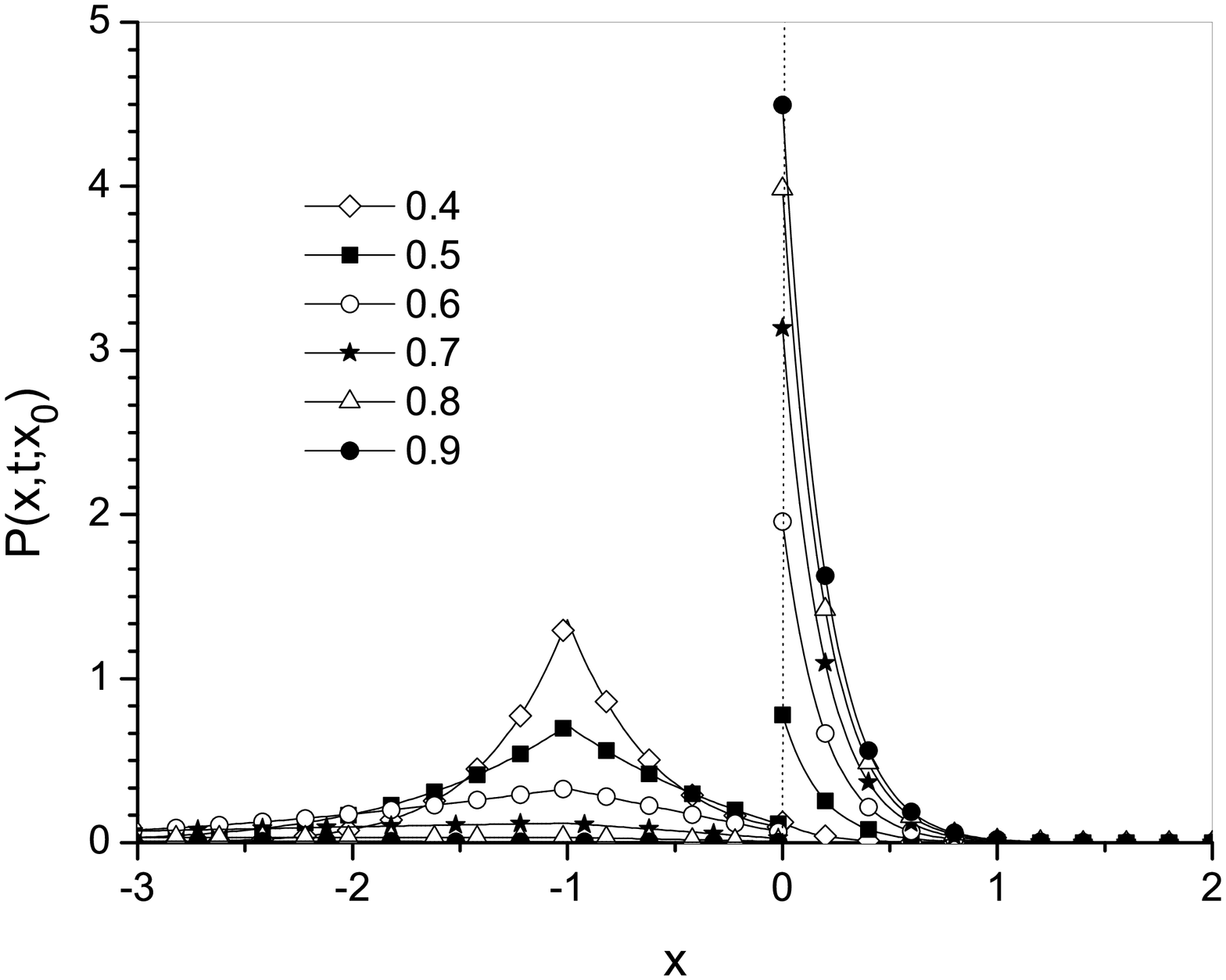}
\caption{The plots ot the Green's functions (\ref{eq59}) and (\ref{eq60}) for various $\alpha_1$ given in the legend, $\alpha_2=0.3$, $t=10^5$, the other parameters are the same as in Fig. \ref{Fig2}.}\label{Fig5}
\end{figure}

In Figs. \ref{Fig2}--\ref{Fig5}, the Green's functions  are presented for the system in which the particle is located in the `faster' medium at the initial moment. The calculations were performed taking 20 first terms in the function Eq. (\ref{eq58}), and in the series in Eq. (\ref{eq60}).

We observe that the concentration in the region $x<x_N$ is similar to the plots occurring in a system with a fully absorbing wall for a sufficiently long time. Namely, in terms of the Laplace transform, the Green's function for the system with a fully reflecting or fully absorbing wall reads 
\begin{equation}\label{eq61}
\hat{P}(x,s;x_0)=\hat{P}_H(x,s;x_0)+\zeta\hat{P}_H(x,s;x_s)\;, 
\end{equation}
where $x_s$ is the point located symmetrically to the point $x_0$ with respect to the wall, $\zeta =-1$ for the system with an absorbing wall, and $\zeta=1$ for the system with a reflecting wall,
\begin{equation}\label{eq62}
\hat{P}_H(x,s;x_0)=\frac{s^{-1+\alpha_1/2}}{2\sqrt{D_1}}{\rm e}^{-\frac{|x-x_0|s^{\alpha_1/2}}{\sqrt{D_1}}}
\end{equation}
is the Laplace transform of the Green's function for a homogeneous system $A$ without a wall. The function (\ref{eq56}) takes the form of the functions for the system with an absorbing wall under the condition $s\ll(2\tilde{\gamma}_1/\tilde{\gamma}_2)^{2/(\alpha_1-\alpha_2)}$. This inequality provides the condition $t\gg (\tilde{\gamma}_2/2\tilde{\gamma}_1\Gamma(1-(\alpha_1-\alpha_2)/2))^{2/(\alpha_1-\alpha_2)}$.

\subsection{From slower to faster region, $\alpha_1<\alpha_2$\label{SecIVB}}

Let $\alpha_1<\alpha_2$ and $\tilde{\gamma}_1\neq\tilde{\gamma}_2$. 
We transform the functions $\tilde{\Lambda}_A(s)$ and $\tilde{\Lambda}_B(s)$, Eqs. (\ref{eq38}) and (\ref{eq39}), into the following form
\begin{equation}\label{eq63}
\tilde{\Lambda}_A(s)=\frac{1-(\tilde{\gamma}_2/\tilde{\gamma}_1)s^{(\alpha_2-\alpha_1)/2}+\tilde{\gamma}_2 s^{\alpha_2/2}}{1+(\tilde{\gamma}_2/\tilde{\gamma}_1)s^{(\alpha_2-\alpha_1)/2}+\tilde{\gamma}_2 s^{\alpha_2/2}}\;,
\end{equation}
\begin{equation}\label{eq64}
\tilde{\Lambda}_B(s)=\frac{2(\tilde{\gamma}_2/\tilde{\gamma}_1)s^{(\alpha_2-\alpha_1)/2}}{1+(\tilde{\gamma}_2/\tilde{\gamma}_1)s^{(\alpha_2-\alpha_1)/2}+\tilde{\gamma}_2 s^{\alpha_2/2}}\;.		
\end{equation}
It is convenient for further calculations to transform the Laplace transform of the boundary condition to the following form, which contains the positive values of the exponent of $s$
\begin{equation}\label{eq65}
\sqrt{\frac{D_2}{D_1}}\frac{\tilde{\gamma}_2}{\tilde{\gamma}_1}s^{\alpha_2-\alpha_1}\hat{P}_A(x_N,s;x_0)
=\hat{P}_B(x_N,s;x_0)+\tilde{\gamma}_2 s^{\alpha_2/2}\hat{P}_B(x_N,s;x_0)\;.
\end{equation}
The inverse Laplace transform of Eq. (\ref{eq65}) reads
\begin{equation}\label{eq66}
\sqrt{\frac{D_2}{D_1}}\frac{\tilde{\gamma}_2}{\tilde{\gamma}_1}\frac{\partial^{\alpha_2-\alpha_1}P_A(x_N,t;x_0)}{\partial t^{\alpha_2-\alpha_1}}
=P_B(x_N,t;x_0)+\tilde{\gamma}_2\frac{\partial^{\alpha_2/2}P_B(x_N,t;x_0)}{\partial t^{\alpha_2/2}}\;.
\end{equation}

As in the previous case, we consider the Laplace transform of the boundary condition and the Green's functions in the limit of small values of $s$. For $s\ll\left(1/\tilde{\gamma}_2\right)^{2/\alpha_2}$ we can omit the last term on the right--hand side of Eq. (\ref{eq65}). Then, for $t\gg \left(\tilde{\gamma}_2/\Gamma(1-\alpha_2/2)\right)^{2/\alpha_2}$ the boundary condition reads
\begin{equation}\label{eq67}
P_B(x_N,t;x_0)=\sqrt{\frac{D_2}{D_1}}\frac{\tilde{\gamma}_2}{\tilde{\gamma}_1}\frac{\partial^{\alpha_2-\alpha_1}P_A(x_N,t;x_0)}{\partial t^{\alpha_2-\alpha_1}}\;.
\end{equation}
We suppose that $s\ll\left(\tilde{\gamma}_1/\tilde{\gamma}_2\right)^{2/(\alpha_2-\alpha_1)}$. So, we have
$\tilde{\Lambda}_A(s)=1-2(\tilde{\gamma}_2/\tilde{\gamma}_1)s^{(\alpha_2-\alpha_1)/2}$ and 
$\tilde{\Lambda}_B(s)=2(\tilde{\gamma}_2/\tilde{\gamma}_1)s^{(\alpha_2-\alpha_1)/2}$.
The above functions and Eqs. (\ref{eq40}) and (\ref{eq41}) provide 
\begin{eqnarray}\label{eq68}
\hat{P}_A(x,s;x_0)=\frac{s^{-1+\alpha_1/2}}{2\sqrt{D_1}}\Bigg[{\rm e}^{-\frac{|x-x_0|s^{\alpha_1/2}}{\sqrt{D_1}}}\\
+\left(1-\frac{2\tilde{\gamma}_2 s^{(\alpha_2-\alpha_1)/2}}{\tilde{\gamma}_1}\right){\rm e}^{-\frac{(2x_N-x-x_0)s^{\alpha_1/2}}{\sqrt{D_1}}}\Bigg]\nonumber\;,
\end{eqnarray}
\begin{equation}\label{eq69}
\hat{P}_B(x,s;x_0)=\frac{\tilde{\gamma}_2 s^{-1+\alpha_2-\alpha_1/2}}{\tilde{\gamma}_1 \sqrt{D_2}}{\rm e}^{-\frac{(x_N-x_0)s^{\alpha_1/2}}{\sqrt{D_1}}}{\rm e}^{-\frac{(x-x_N)s^{\alpha_2/2}}{\sqrt{D_2}}}\;.
\end{equation}
From (\ref{eq58}), (\ref{eq68}) and (\ref{eq69}) we get for $t\gg {\rm max}[t_1,t_2]$, where $t_1=(\tilde{\gamma}_2/\Gamma(1-\alpha_2/2))^{2/\alpha_2}$, $t_2=\left(\tilde{\gamma}_2/\tilde{\gamma}_1\Gamma(1-(\alpha_2-\alpha_1)/2)\right)^{2/(\alpha_2-\alpha_1)}$
\begin{eqnarray}\label{eq70}
P_A(x,t;x_0)=\frac{1}{2\sqrt{D_1}}\Bigg[f_{-1+\alpha_1/2,\alpha_1/2}\left(t;\frac{|x-x_0|}{\sqrt{D_1}}\right)\\
+f_{-1+\alpha_1/2,\alpha_1/2}\left(t;\frac{2x_N-x-x_0}{\sqrt{D_1}}\right)\Bigg]\nonumber\\
-\frac{\tilde{\gamma}_2}{\sqrt{D_1}\tilde{\gamma}_1}f_{-1+\alpha_2,\alpha_1/2}\left(t;\frac{2x_N-x-x_0}{\sqrt{D_1}}\right)\;,\nonumber
\end{eqnarray}
\begin{eqnarray}\label{eq71}
P_B(x,t;x_0)=\frac{\tilde{\gamma}_2}{\sqrt{D_2}\tilde{\gamma}_1}\sum_{n=0}^\infty\frac{1}{n!}\left(\frac{x_N-x}{\sqrt{D_2}}\right)^n \\
\times f_{-1-\alpha_1/2+\alpha_2(1+n/2),\alpha_1/2}\left(t;\frac{x_N-x_0}{\sqrt{D_1}}\right)\;.\nonumber
\end{eqnarray}

\begin{figure}[!ht]
\centering
\includegraphics[height=6.0cm]{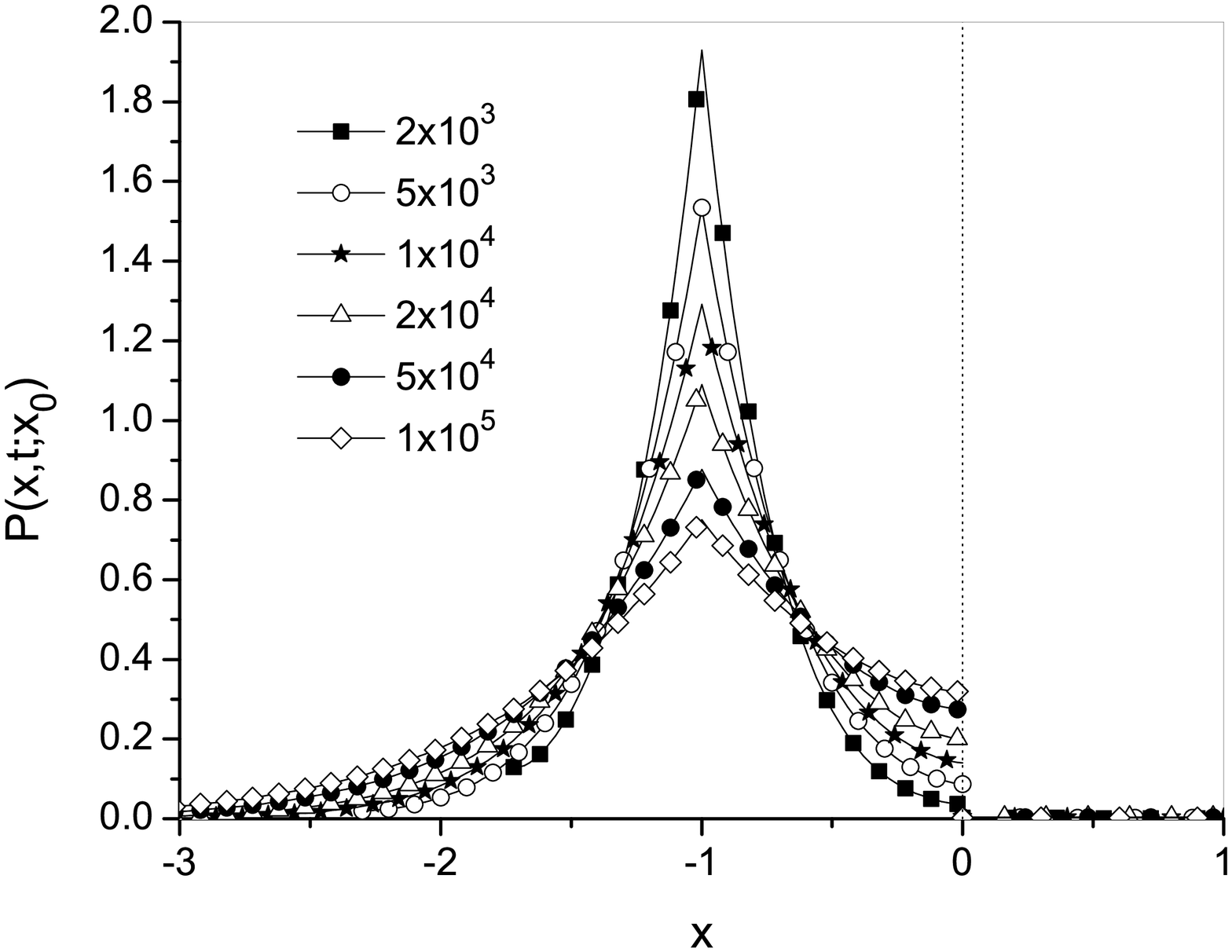}
\caption{The plots of the Green's functions (\ref{eq70}) and (\ref{eq71}) for various time given in the legend, $\alpha_1=0.5$, $\alpha_2=0.9$, the other parameters are the same as in Fig. \ref{Fig2}.}\label{Fig6}
\end{figure}

\begin{figure}[!ht]
\centering
\includegraphics[height=6.0cm]{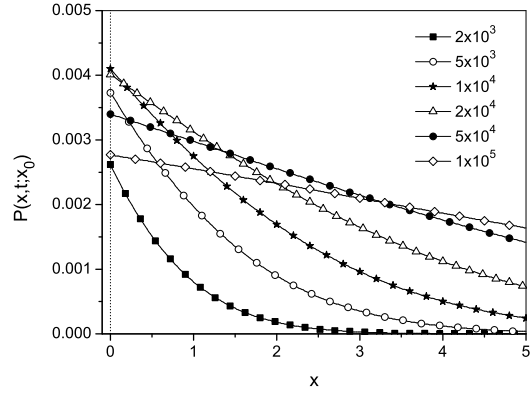}
\caption{The fragment of Fig. \ref{Fig6} presented in the other scale of space variable.}\label{Fig7}
\end{figure}

\begin{figure}[!ht]
\centering
\includegraphics[height=6.0cm]{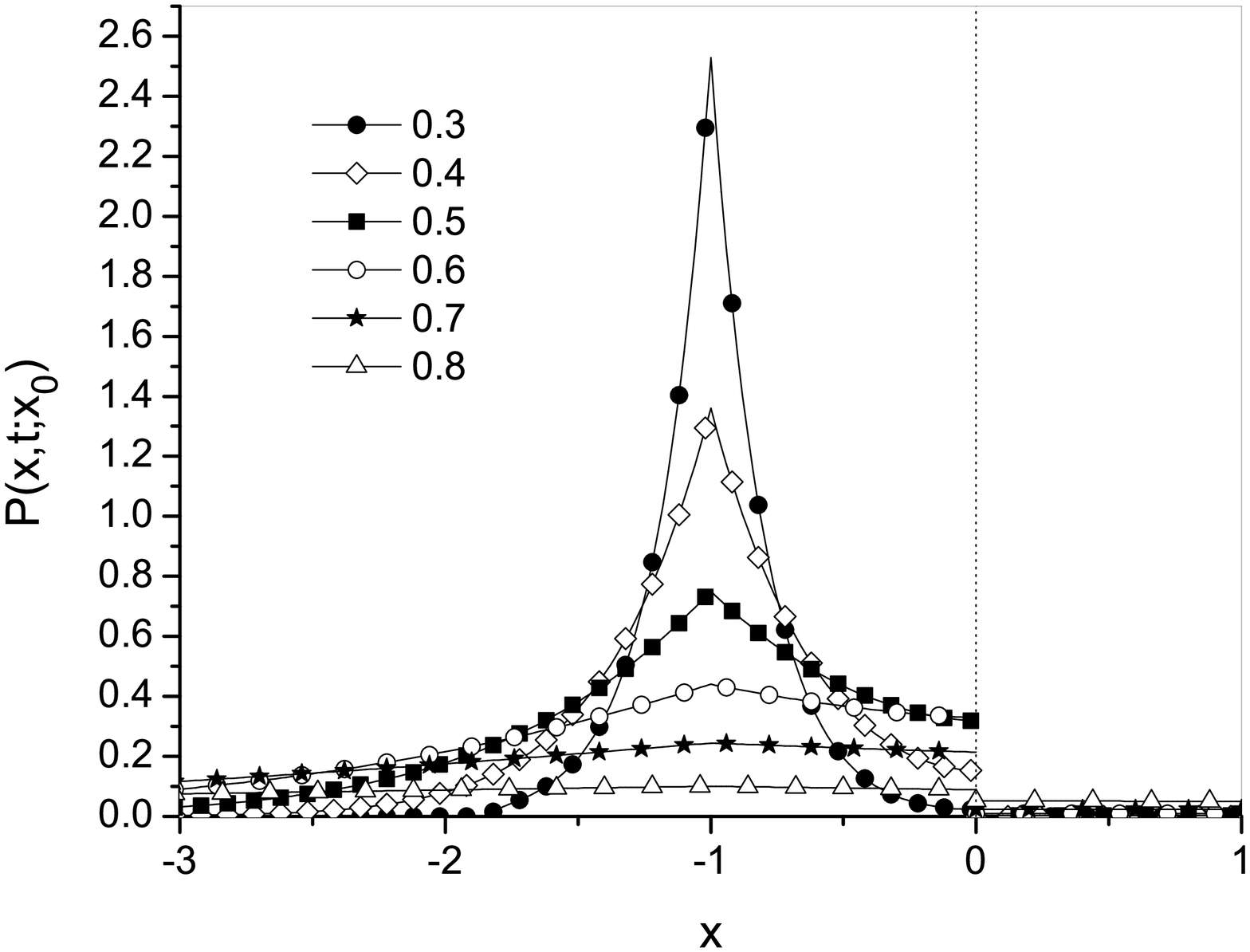}
\caption{The plots ot the Green's functions (\ref{eq70}) and (\ref{eq71}) for various $\alpha_1$ which are given in the legend, $\alpha_2=0.9$, $t=10^5$, the other parameters are the same as in Fig. \ref{Fig2}.}\label{Fig8}
\end{figure}

We observe that the concentration in the region $x<x_N$ is similar to the plots occurring in a system with a fully reflecting wall for sufficiently long times. The function (\ref{eq68}) takes the form of the functions for the system with a reflecting wall under condition $s\ll(2\tilde{\gamma}_2/\tilde{\gamma}_1)^{2/(\alpha_2-\alpha_1)}$. This equality provides the condition $t\gg (\tilde{\gamma}_1/2\tilde{\gamma}_2\Gamma(1-(\alpha_2-\alpha_1)/2))^{2/(\alpha_2-\alpha_1)}$.

\subsection{The case of $\alpha_1=\alpha_2$\label{SecIVC}}

Let $\alpha_1=\alpha_2=\alpha$. 
The functions $\tilde{\Lambda}_A(s)$ and $\tilde{\Lambda}_B(s)$ take the following form
\begin{equation}\label{eq72}
\tilde{\Lambda}_A(s)=\frac{\tilde{\gamma}_1-\tilde{\gamma}_2+\tilde{\gamma}_1\tilde{\gamma}_2 s^{\alpha/2}}{\tilde{\gamma}_1+\tilde{\gamma}_2+\tilde{\gamma}_1\tilde{\gamma}_2 s^{\alpha/2}}\;,
\end{equation}
\begin{equation}\label{eq73}
\tilde{\Lambda}_B(s)=\frac{2}{\tilde{\gamma}_1+\tilde{\gamma}_2+\tilde{\gamma}_1\tilde{\gamma}_2 s^{\alpha/2}}\;.		
\end{equation}
In terms of the Laplace transform the boundary condition Eq. (\ref{eq42}) is the following
\begin{eqnarray}\label{eq74}
\hat{P}_A(x_N,s;x_0)
=\sqrt{\frac{D_2}{D_1}}\left[\frac{\tilde{\gamma}_1}{\tilde{\gamma}_2}+\tilde{\gamma}_1 s^{\alpha/2}\right]\hat{P}_B(x_N,s;x_0)\;.
\end{eqnarray}
The inverse Laplace transform of Eq. (\ref{eq74}) reads
\begin{equation}\label{eq75}
P_A(x_N,t;x_0)
=\sqrt{\frac{D_2}{D_1}}\frac{\tilde{\gamma}_1}{\tilde{\gamma}_2}P_B(x_N,t;x_0)
+\sqrt{\frac{D_2}{D_1}}\tilde{\gamma}_1\frac{\partial^{\alpha/2}P_B(x_N,t;x_0)}{\partial t^{\alpha/2}}\;.
\end{equation}
In the limit of small $s$, $s\ll\left(1/\tilde{\gamma}_2\right)^{2/\alpha}$, the Laplace transform of boundary condition (\ref{eq74}) reads
\begin{equation}\label{eq76}
\hat{P}_A(x_N,s;x_0)
=\sqrt{\frac{D_2}{D_1}}\frac{\tilde{\gamma}_1}{\tilde{\gamma}_2}\hat{P}_B(x_N,s;x_0)\;.
\end{equation}
The inverse Laplace transform of Eq. (\ref{eq76}) is the following
\begin{equation}\label{eq77}
P_A(x_N,t;x_0)
=\sqrt{\frac{D_2}{D_1}}\frac{\tilde{\gamma}_1}{\tilde{\gamma}_2}P_B(x_N,t;x_0)\;.
\end{equation}
Assuming additionally: $s\ll\left(\tilde{\gamma}_1\tilde{\gamma}_2/(\tilde{\gamma}_1+\tilde{\gamma}_2)\right)^{2/\alpha}$,
from Eqs. (\ref{eq72}) and (\ref{eq73}) we obtain
\begin{equation}\label{eq78}
\tilde{\Lambda}_A(s)=\tilde{\kappa}_1+\tilde{\kappa}_2 s^{\alpha/2}\;,
\end{equation}
\begin{equation}\label{eq79}
\tilde{\Lambda}_B(s)=1-\tilde{\kappa}_1-\tilde{\kappa}_2 s^{\alpha/2}\;,
\end{equation}
where
\begin{equation}\label{eq80}
\tilde{\kappa}_1=\frac{\tilde{\gamma}_1-\tilde{\gamma}_2}{\tilde{\gamma}_1+\tilde{\gamma}_2}\;,\;\tilde{\kappa}_2=\frac{2\tilde{\gamma}_1 \tilde{\gamma}^2_2}{(\tilde{\gamma}_1+\tilde{\gamma}_2)^2}\;.
\end{equation}
The inverse formulae to Eq. (\ref{eq80}) are $\tilde{\gamma}_1=2\tilde{\kappa}_2/(1-\tilde{\kappa}_1)^2$ and $\tilde{\gamma}_2=2\tilde{\kappa}_2/(1-\tilde{\kappa}_2^2)$.
Taking into account Eqs. (\ref{eq78}) and (\ref{eq79}) we obtain
\begin{equation}\label{eq82}
\hat{P}_A(x,t;x_0)=\frac{s^{-1+\alpha/2}}{2\sqrt{D_1}}\Bigg[{\rm e}^{-\frac{|x-x_0|s^{\alpha/2}}{\sqrt{D_1}}}
+\left(\tilde{\kappa}_1+\tilde{\kappa}_2 s^{\alpha/2}\right){\rm e}^{-\frac{(2x_N-x-x_0)s^{\alpha/2}}{\sqrt{D_1}}}\Bigg]\;,
\end{equation}
\begin{equation}\label{eq83}
\hat{P}_B(x,t;x_0)=\frac{s^{-1+\alpha/2}}{2\sqrt{D_2}}\left(1-\tilde{\kappa}_1-\tilde{\kappa}_2 s^{\alpha/2}\right)
{\rm e}^{-\frac{(x_N-x_0)s^{\alpha/2}}{\sqrt{D_1}}}{\rm e}^{-\frac{(x-x_N)s^{\alpha/2}}{\sqrt{D_2}}}\;.
\end{equation}

\begin{figure}[!ht]
\centering
\includegraphics[height=6.0cm]{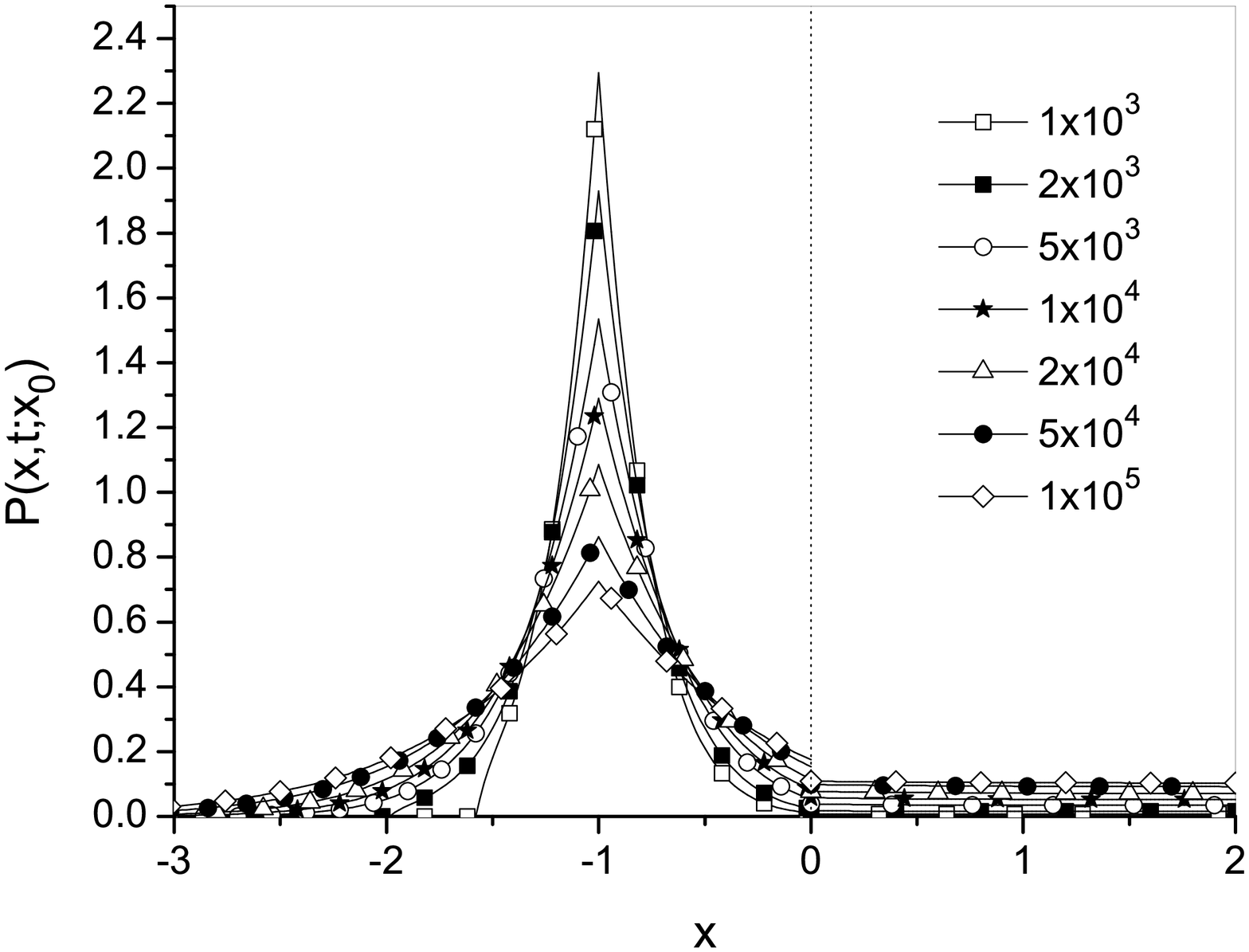}
\caption{The plots ot the Green's functions (\ref{eq84}) and (\ref{eq85}) for various times which are given in the legend, $\alpha_1=\alpha_2=0.5$, $D_1=0.001$, $D_2=0.002$, $\tilde{\gamma}_1=0.06$, $\tilde{\gamma}_2=0.05$, $x_0=-1$, and $x_N=0$.}\label{Fig9}
\end{figure}

The inverse Laplace transform of Eqs. (\ref{eq82}) and (\ref{eq83}) reads for $t\gg\left((\tilde{\gamma}_1+\tilde{\gamma}_2)/\tilde{\gamma}_1\tilde{\gamma}_2\Gamma(1-\alpha/2)\right)^{2/\alpha}$
\begin{eqnarray}\label{eq84}
P_A(x,t;x_0)=\frac{1}{2\sqrt{D_1}}\Bigg[f_{-1+\alpha/2,\alpha/2}\left(t;\frac{|x-x_0|}{\sqrt{D_1}}\right)\\
+\tilde{\kappa}_1 f_{-1+\alpha/2,\alpha/2}\left(t;\frac{2x_N-x-x_0}{\sqrt{D_1}}\right)\nonumber\\
+\tilde{\kappa}_2f_{-1+\alpha,\alpha/2}\left(t;\frac{2x_N-x-x_0}{\sqrt{D_1}}\right)\Bigg]\;,\nonumber
\end{eqnarray}
\begin{eqnarray}\label{eq85}
P_B(x,t;x_0)=\frac{1}{2\sqrt{D_2}}\sum_{n=0}^\infty\frac{1}{n!}\left(\frac{x_N-x}{\sqrt{D_2}}\right)^n \\
\times \Big[(1-\tilde{\kappa}_1)f_{-1+\alpha(1+n/2),\alpha_1/2}\left(t;\frac{x_N-x_0}{\sqrt{D_1}}\right)\nonumber\\
-\tilde{\kappa}_2 f_{-1+\alpha(3+n)/2,\alpha/2}\left(t;\frac{x_N-x_0}{\sqrt{D_1}}\right)\Big]\nonumber\;.
\end{eqnarray}
In this case the plots of Green's functions are qualitatively similar to the ones obtained for the system in which a membrane is located in a homogeneous medium, see \cite{tk}. In contrast to the case of $\alpha_1\neq\alpha_2$, we do not observe any Green's function characteristic of a system with an absorbing or reflecting wall.

\section{Final remarks\label{SecV}}

The main results presented in this paper are both the Green's functions, Eqs. (\ref{eq40}) and (\ref{eq41}), and boundary conditions at a thin membrane, Eqs. (\ref{eq42}) and (\ref{eq48}), given in terms of the Laplace transform. From these equations we can derive both Green's functions and boundary conditions at the membrane in time domain, over a limit of long time, for various relations between the subdiffusion parameters $\alpha_1$ and $\alpha_2$. The examples for the calculation of the Green's functions and the boundary conditions from their Laplace transforms are shown in Sec. \ref{SecIV}. The boundary conditions at a thin membrane can be used to calculate particles' concentration profiles in systems consisting of many parts separated by thin membranes, including the case of subdiffusion in a system with a thick membrane.

In this paper we also present the new method of deriving the Green's function for a system in which two different media are separated by a thin partially permeable membrane. Within this method we first consider a particle's random walk in a system with a discrete time and space variable in which the particle may vanish due to reactions with constant probabilities defined separately for each medium. The reactions fulfil the supporting role. Since we link the reaction parameters to subdiffusion parameters, the presence of reactions allow us to include the subdiffusion parameters of both media into the Green's functions derived from a random walk model. So, we move from discrete to continuous variables in the obtained formulae and eliminate the reactions.  Finally, we obtained the Green's functions for the system without reactions. The functions obtained were used to determine the boundary conditions at the thin membrane. The results presented in this paper are also valid when normal diffusion occurs in the system. We obtain the Green's functions and boundary conditions for a normal diffusion case by placing $\alpha_1=1$ and/or $\alpha_2=1$ into the obtained formulae. 

The model based on the random walk model with a discrete time and space variable seems to be oversimplified. However, it provides somewhat nontrivial results which have a simple physical interpretation. We add that such models were used to model subdiffusion processes in membrane systems \cite{tk,tk1}, as was mentionad earlier in this paper, as well as in the system with reactions \cite{kl2014,koszt2014}.

The Green's functions differ qualitatively depending on whether a particle starts its subdiffusive random walk from a faster or slower medium, which is shown in Figs. \ref{Fig2}--\ref{Fig7}. If the particle starts from a `faster' medium, then its random walk in this medium is performed just as in the medium with an absorbing membrane, whereas when the particle starts from a `slower' medium, then its random walk in this medium is similar to subdiffusion occurring in a system with a reflecting membrane over a long time limit. In both cases, the probability of finding the particle in the `slower' medium significantly increases over time.

We have considered the process in which $x_0< x_N$. The Green's functions obtained in this paper can be easily transformed for the case of $x_0> x_N$ using the symmetry argument, which, in practice, means that the following conversion is made: $\alpha_1\longrightarrow\alpha_2$, $\alpha_2\longrightarrow\alpha_1$, $D_1\longrightarrow D_2$, $D_2\longrightarrow D_1$, $\tilde{\gamma}_1\longrightarrow\tilde{\gamma}_2$, $\tilde{\gamma}_2\longrightarrow\tilde{\gamma}_1$, $x-x_0\longrightarrow x_0-x$, $x-x_N\longrightarrow x_N-x$, and $x_N-x_0\longrightarrow x_0-x_N$. If we consider a system with many particles and if the particles move independently of each other, the concentration of particles $C(x,t)$ can be calculated by means of the following formula $C(x,t)=\int_{-\infty}^\infty P(x,t;x_0)C(x_0,0)dx_0$.  The procedure of finding the concentration profiles for the arbitrarily chosen initial condition $C(x_0,0)$ is identical to the procedure described in \cite{tk} (see Eqs. (78)--(90) in the above cited paper), where subdiffusion in a system in which a thin membrane separates two identical media is taken into account.

The results presented in this paper agree in part with the results obtained in \cite{korabel1,korabel2,korabel3}, in which subdiffusion in a system which consists of two different subdiffusive media are joined at point $x_N$. However, in these papers, only the Green's function for the case of $x_0=x_N$ was found. The Laplace transforms of the Green's functions derived in the above cited paper coincides with Eqs. (\ref{eq56}), (\ref{eq57}) and Eqs. (\ref{eq68}), (\ref{eq69}) presented in this paper, only in the case of $x_0=x_N$. The boundary condition derived in \cite{korabel3} coincides with boundary conditions Eqs. (\ref{eq55}) and (\ref{eq67}) in this paper which are the approximation of more general boundary conditions, Eqs. (\ref{eq53}) and (\ref{eq66}), respectively, over a long time limit. The model presented in this paper allows us to consider the subdiffusion of a particle initially located at an arbitrarily chosen point $x_0$.

Below, we add the comments concerning the technical aspects of the calculations. The moving from a discrete to continuous spatial variable is interpreted slightly differently from mathematical or physical points of view. Mathematically, this moving consists of calculating the limit $\epsilon \longrightarrow 0$ in the Green's functions already obtained for a discrete spatial variable. Physically, supposing that $\epsilon$ is very small but finite, one may expand the Green's functions in the series with respect to $\epsilon$ and then omit the terms of the order equal to or larger than 1 with respect to $\epsilon$. The difference between both approaches is caused by the fact that in the limit $\epsilon\longrightarrow 0$, the probability distribution $\omega(t)$ of the waiting time for a particle to take its next step loses its physical interpretation; this problem is briefly discussed just below Eq. (\ref{eq34}). However, these two approaches lead to the same result; so, in practice, it is possible to use both in order to find the functions for a continuous spatial variable.

One of the criteria that provides the utility of the used approximation is that the Green's functions obtained for a membrane system should depend on the parameters of membrane permeability including the case of $\tilde{\gamma}_1=\tilde{\gamma}_2$ (see the discussion presented in \cite{tk}).
However, when $\alpha_1\neq\alpha_2$, several leading terms in the series which approximate the functions (\ref{eq38}) and (\ref{eq39}) with respect to $s$, are dependent on the ratio $\tilde{\gamma}_1/\tilde{\gamma}_2$ alone; the number of leading terms increases when $|\alpha_1-\alpha_2|$ becomes smaller. When $\tilde{\gamma}_1=\tilde{\gamma}_2$, in order to include the membrane permeability parameters in the approximate Green's functions, we have to include a relatively large number of terms, especially when $\alpha_1\neq\alpha_2$ and $|\alpha_1-\alpha_2|\ll 1$. We note that this problem does not occur in the system with $\alpha_1=\alpha_2$. 
In this paper, we consider the subdiffusion in the composite membrane system for the case of $\alpha_1\neq\alpha_2$, assuming $\tilde{\gamma}_1\neq\tilde{\gamma}_2$. We presume that the last condition is not significantly limiting in our considerations. Although from the mathematical point of view, the cases $\tilde{\gamma}_1=\tilde{\gamma}_2$ and $\tilde{\gamma}_1\neq\tilde{\gamma}_2$ should be treated `equitably', from the physical point of view, the case of $\tilde{\gamma}_1\neq\tilde{\gamma}_2$ is preferable. The case of $\tilde{\gamma}_1=\tilde{\gamma}_2$ naturally occurs when a symmetrical thin membrane is located in a homogeneous system in which $\alpha_1=\alpha_2$. The Green's functions for such a system are a special case of the functions obtained in Sec. IIIC. For the system in which $\alpha_1\neq\alpha_2$, the coefficients $\tilde{\gamma}_1=\gamma_1/\sqrt{D_1}$ and $\tilde{\gamma}_2=\gamma_2/\sqrt{D_2}$, which are given in different physical units, are equal for a very particular combination of parameters which reads
$D_1/D_2=\left(\gamma_1/\gamma_2\right)^2$. 
We note that $\gamma_i$ is measured in the units $m$, whereas $D_i$ is measured in units $m^2/s^{\alpha_i}$, $i=1,2$. Changing the units of time provides a change on the left--hand side of the above relation, with its right--hand side remaining unchanged. Thus, the assumption $\tilde{\gamma}_1\neq\tilde{\gamma}_2$ seems to have a good basis for the case of $\alpha_1\neq\alpha_2$.

\section*{Acknowledgments}

The author would like to thank Dr. Katarzyna D. Lewandowska for her helpful discussions.
This paper was partially supported by the Polish National Science Centre under grant No. 2014/13/D/ST2/03608.

\section*{Appendix I}

We show the method of calculating the functions Eqs. (\ref{eq10})--(\ref{eq14}). Using Eqs. (\ref{eq4})--(\ref{eq9}), we obtain
\begin{eqnarray}
  \label{a1}\frac{1}{z}\left[S_A(m,z;m_0)-\delta_{m,m_0}\right]=\frac{1}{2}S_A(m-1,z;m_0)+\frac{1}{2}S_A(m+1,z;m_0)\nonumber
	\\-R_1 S_A(m,z;m_0)  \qquad m\neq N, N+1\;,\\
     \nonumber\\
  \label{a2}\frac{1}{z}\left[S_A(N,z;m_0)-\delta_{N,m_0}\right]=\frac{1}{2}S_A(N-1,z;m_0)+\frac{1-q_2}{2}S_B(N+1,z;m_0)\nonumber
  \\+\frac{1-q_1}{2}S_A(N+1,z;m_0)-R_1 S_A(N,z;m_0)\;,\\
     \nonumber\\
  \label{a3}\frac{1}{z}\left[S_B(N+1,z;m_0)-\delta_{N+1,m_0}\right]=\frac{1-q_1}{2}S_A(N,z;m_0)+\frac{1}{2}S_B(N+2,z;m_0)\nonumber
  \\+\frac{q_2}{2}S_B(N+1,z;m_0)-R_2 S_B(N+1,z;m_0)\;,\\
    \nonumber\\
   \label{a4}\frac{1}{z}\left[S_B(m,z;m_0)-\delta_{m,m_0}\right]=\frac{1}{2}S_B(m-1,z;m_0)+\frac{1}{2}S_B(m+1,z;m_0)\nonumber
	\\-R_2 S_B(m,z;m_0)\;,  \qquad m>N+1\;. 
\end{eqnarray} 
In order to solve Eqs. (\ref{a1})--(\ref{a4}), we use the following generating functions with respect to the space variable
\begin{eqnarray}
G_A(u,z;m_0)=\sum_{m=-\infty}^N u^m S_A(m,z;m_0)\;,\label{a5}\\
G_B(u,z;m_0)=\sum_{m=N+1}^\infty u^m S_B(m,z;m_0)\;.\label{a6}
\end{eqnarray}
After calculating, we obtain
\begin{eqnarray}\label{a7}
G_A(u,z;m_0)\left[1-\frac{z}{2}\left(u+\frac{1}{u}+R_1 z\right)\right]\\
=\frac{z}{2}u^N\left[q_1 S_A(N,z;m_0)+(1-q_2)S_B(N+1,z;m_0)\right]\nonumber\\
+u^{m_0}-\frac{z}{2}u^{N+1}S_A(N,z;m_0)\nonumber\;,
\end{eqnarray}

\begin{eqnarray}\label{a8}
G_B(u,z;m_0)\left[1-\frac{z}{2}\left(u+\frac{1}{u}+R_2 z\right)\right]\\
=\frac{z}{2}u^{N+1}\left[(1-q_1) S_A(N,z;m_0)+q_2 S_B(N+1,z;m_0)\right]\nonumber\\
-\frac{z}{2}u^{N}S_B(N+1,z;m_0)\nonumber\;.
\end{eqnarray}
The generating function $S_i$ can be obtained by means of the following formula
\begin{equation}\label{a9}
S_i(m,z;m_0)=\frac{1}{2\pi i}\oint_{K({\bf 0},1)}\frac{G_i(u,z;m_0)}{u^{m+1}}du\;,
\end{equation}
where $i=A,B$, the integration is carried out along the unit circle $K$ centered at the point ${\bf 0}=(0,0)$ to be consistent with the increasing argument of a complex number. 
Using the integral formula
\begin{eqnarray}\label{a10}
\frac{1}{2\pi i}\oint_{K({\bf 0},1)}\frac{u^{m_0}}{u^{m+1}\left[1-\frac{z}{2}\left(u+\frac{1}{u}\right)+zR\right]}du\\
=\left(\frac{1+zR-\sqrt{(1+zR)^2-z^2}}{z}\right)^{|m-m_0|}\frac{1}{\sqrt{(1+zR)^2-z^2}}\nonumber\;,
\end{eqnarray}
from Eqs. (\ref{a7})--(\ref{a9}), we obtain, after calculations, Eqs. (\ref{eq10})--(\ref{eq14}).

\section*{Appendix II}

We suppose that
\begin{equation}\label{a11}
q_1(\epsilon)=1-\frac{f_1(\epsilon)}{\gamma_1}\;,\;q_2(\epsilon)=1-\frac{f_2(\epsilon)}{\gamma_2}\;,
\end{equation}
where $f_1(\epsilon)$, $f_2(\epsilon)$ are positive functions as yet to be determined, and $\gamma_1$ and $\gamma_2$ are reflection membrane coefficients defined for the continuous system.
From Eqs. (\ref{eq28}), (\ref{eq29}), (\ref{eq35}) and (\ref{a11}) we obtain
\begin{equation}\label{a12}
\tilde{\Lambda}_A(s)=\frac{\frac{1}{\epsilon}\left[\frac{f_2(\epsilon)s^{\alpha_1/2}}{\gamma_2\sqrt{D_1}}-\frac{f_1(\epsilon)s^{\alpha_2/2}}{\gamma_1\sqrt{D_2}}\right]+\frac{s^{(\alpha_1+\alpha_2)/2}}{\sqrt{D_1 D_2}}}{\frac{1}{\epsilon}\left[\frac{f_2(\epsilon)s^{\alpha_1/2}}{\gamma_2\sqrt{D_1}}+\frac{f_1(\epsilon)s^{\alpha_2/2}}{\gamma_1\sqrt{D_2}}\right]+\frac{s^{(\alpha_1+\alpha_2)/2}}{D_1 D_2}}\;,
\end{equation}
\begin{equation}\label{a13}
\tilde{\Lambda}_B(s)=\frac{\frac{2}{\epsilon}\frac{f_1(\epsilon)s^{\alpha_1/2}}{\gamma_1\sqrt{D_2}}}{\frac{1}{\epsilon}\left[\frac{f_2(\epsilon)s^{\alpha_1/2}}{\gamma_2\sqrt{D_1}}+\frac{f_1(\epsilon)s^{\alpha_2/2}}{\gamma_1\sqrt{D_2}}\right]+\frac{s^{(\alpha_1+\alpha_2)/2}}{D_1 D_2}}\;.
\end{equation}
The only form of the functions $f_1(\epsilon)$ and $f_2(\epsilon)$ which ensures that the functions $\tilde{\Lambda}_A(s)$ and $\tilde{\Lambda}_B(s)$, Eqs. (\ref{a12}) and (\ref{a13}), are finite and depend on membrane's reflection parameters for any $\gamma_1$ and $\gamma_2$ within the limit $\epsilon\longrightarrow 0$ is
\begin{equation}\label{a14}
f_1(\epsilon)\equiv f_2(\epsilon)\equiv \epsilon\;.
\end{equation}

\end{document}